\documentstyle[prd,aps,preprint,tighten,epsfig]{revtex}

\begin{document}

\draft

\title{Neutrino Masses and Lepton-flavor-violating $\tau$ Decays in the Supersymmetric Left-right Model}

\author{{\bf Wei Chao} \thanks{E-mail: chaowei@mail.ihep.ac.cn}}
\address{
Institute of High Energy Physics, Chinese Academy of Sciences, \\
P.O. Box 918, Beijing 100049, China} \maketitle
\begin{abstract}
In the supersymmetric left-right model, the light neutrino masses
are given by the Type-II seesaw mechanism. A duality property
about this mechanism indicates that there exist eight possible
Higgs triplet Yukawa couplings which result in the same neutrino
mass matrix. In this paper, We work out the one-loop
renormalization group equations for the effective neutrino mass
matrix in the supersymmetric left-right model. The stability of
the Type-II seesaw scenario is briefly discussed. We also study
the lepton-flavor-violating processes ( $\tau\rightarrow
\mu\gamma$ and $\tau\rightarrow e\gamma$ ) by using the
reconstructed Higgs triplet Yukawa couplings.
\end{abstract}

\newpage
\section{introduction}
Current solar, atmospheric, reactor and accelerator neutrino
oscillation experiments have provided us with very convincing
evidence that neutrinos have non-vanishing masses and lepton
flavors are mixed\cite{exp1,exp2,exp3,exp4,exp5}. A global
analysis of current experimental data yields $30^\circ \leq
\theta_{12} \leq 38^\circ$, $36^\circ \leq \theta_{23} \leq
54^\circ$ and $0^{} < \theta_{13} < 10^\circ$ as well as $\Delta
m^2_{21} \equiv m^2_2 - m^2_1 = (7.2 \cdot\cdot\cdot 8.9) \times
10^{-5} ~{\rm eV}^2$ and $\Delta m^2_{32} \equiv m^2_3 - m^2_2 =
\pm (2.1 \cdot\cdot\cdot 3.1) \times 10^{-3} ~{\rm eV}^2$ at the
$99\%$ confidence level\cite{Vissani}, but three CP-violating
phases (i.e., the Dirac phase $\delta$ and the Majorana phases
$\rho$ and $\sigma$ ) are entirely unrestricted. These important
results indicate that there should be a more fundamental theory
beyond the standard model, in which three neutrinos are massless
Weyl particles. One possible candidate for such a theory is the
supersymmetric version of the left-right symmetric
model\cite{LR0}, which provides a natural embedding of the seesaw
mechanism for small neutrino masses\cite{Ji}.

The supersymmetric left-right model\cite{lrsusy1,lrsusy2} is based
on the gauge group $SU(3)_C^{}\times SU(2)_{L}^{}\times
SU(2)_R^{}\times U(1)_{B-L}$. The quarks and leptons transform
under the gauge group as $Q(3, 2,1,1/3)$, $Q^c(3^*, 1, 2, -1/3)$,
$L(1, 2, 1, -1)$ and $L^c(1, 1, 2, 1)$.  In the gauge sector,
there are triplet gauge bosons $(W^+,W^-,W^0)_L^{}$,
$(W^+,W^-,W^0)_R^{}$ corresponding to $SU(2)_L^{}$ and
$SU(2)_R^{}$ and a vector boson $V$ corresponding to
$U(1)_{B-L}^{}$, together with their superpartners. Fermion masses
arise from the Yukawa  coupling between quarks, leptons and Higgs
bi-doublets: $\Phi_u^{}(2, 2, 0)$ and $\Phi_d^{}(2, 2, 0)$. The
gauge group $SU(2)_R^{}\times U(1)_{B-L}^{}$ is broken to the
hypercharge symmetry $U(1)_Y^{}$ by the vacuum expectation value
(vev) of a $B-L=-2$ Higgs triplet $\Delta^c(1, 1, 3, -2)$ which is
accompanied by a left-handed Higgs triplet $\Delta(1, 3, 1, 2)$.
The choice of the triplets is preferred because with this choice
the seesaw arises from purely renormalizable interactions. In
addition to $\Delta\ {\rm and}\ \Delta^c$, the model must contain
their conjugate fields $\bar{\Delta}\ {\rm and}\ \bar{\Delta}^c$
to insure the cancellation of the anomalies that would otherwise
occur in the fermionic sector. Given their strange quantum
numbers, the $\bar{\Delta}$ and $\bar{\Delta}^c$ do not couple to
any of the particles in the theory, and thus their contributions
are negligible for any phenomenological studies. The gauge
invariant part of the matter superpotential can be written as
\begin{eqnarray}
W=Y_q^{i}(Q^c)^T\tilde{\Phi}_i Q+Y_l^{i}(L^c)^T\tilde{\Phi}_i
L+i\left({\cal F}L^T\tau_2^{}\Delta L+{\cal F}_c
{L^c}^T\tau_2^{}\Delta^c L^c\right) ,
\end{eqnarray}
where $\tilde{\Phi}_i^{}=i\tau_2^{} \Phi_i^{} $ is defined and
$i=u, d$. All the couplings $Y_q^i$, $Y_l^i$, ${\cal F}$ and
${\cal F}_c$ are complex with ${\cal F}$ and ${\cal F}_c$ being
symmetric matrices. The left-right symmetry implies $Y_\alpha^i=
(Y_\alpha^i)^\dagger$ $(\alpha=q, l)$ and ${\cal F}= {\cal
F}_c^*$. Given the vevs of $\Phi_{u, d}$, $\Delta$ and $\Delta^c$,
\begin{eqnarray}
\langle\Phi_u\rangle&=&\left(\matrix{\kappa_u&0\cr0&0\cr}\right),\hspace{1.5cm}
\langle\Phi_d\rangle=\left(\matrix{0&0\cr0&\kappa_d\cr}\right),
\nonumber\\ \langle\Delta\rangle&=&\left(\matrix{0&0\cr
v_L&0\cr}\right),\hspace{1.5cm}
\langle\Delta^c\rangle=\left(\matrix{0&0\cr v_R&0\cr}\right),
\end{eqnarray}
the gauge group is broken to $U(1)_{em}^{}$ and the up-type quark,
down-type quark, charged lepton and Dirac neutrino mass matrices
turn out to be: $M_u^{}=Y_q^{u}\kappa_u^{}$,
$M_d^{}=Y_q^{d}\kappa_d^{}$, $M_l^{}=Y_l^{u}\kappa_d^{}$ and
$M_D^{}=Y_l^{d}\kappa_u^{}$. Meanwhile, the left- and right-handed
Majorana neutrino mass matrices can be obtained from the
corresponding mass terms in Eq. (1) once the Higgs triplets
$\Delta^{}$ and $\Delta^{c}$ acquire their vevs: $M^{}_L \simeq
v^{}_L {\cal F}$ and $M^{}_R \simeq v^{}_R {\cal F}$. Integrating
out the heavy particles (i.e., the right-handed Majorana neutrinos
and Higgs triplet), one obtains the effective mass matrix for
three light (left-handed) Majorana neutrinos via the Type-II
seesaw mechanism\cite{origin0}:
\begin{eqnarray}
M_\nu^{} \; \simeq \; M_L^{} - M_D^{T} M_R^{-1} M_D^{} \; \simeq
\; v_L^{} {\cal F} - {1\over v_R^{}} M_D^{T} {\cal F}^{-1} M_D^{}
\; .
\end{eqnarray}
We may find that the same coupling ${\cal F}$ appears in both
contributions just because of the left-right symmetry.

Note that Eq. (3) has a duality property\cite{first}: given
$M_D^{}$, there exist eight possible Higgs triplet Yukawa
couplings which result in the same neutrino mass matrix. The
stability of the duality relation and some other phenomena based
on this have been investigated recently. In this paper we perform
a full analysis of the renormalization group equations (RGEs) of
the effective neutrino mass operators. We write down the
$\beta$-functions of the effective neutrino mass operators and
discuss the stability of the Type-II seesaw mechanism.
Lepton-flavor-violating decays in the supersymmetric left-right
model are different from that in the minimal supersymmetric
standard model (MSSM) for the existence of the Higgs triplet
Yukawa coupling ${\cal F}$\cite{theotrical,lfvlr}. In this paper,
we calculate the ${\rm BR}(\tau\rightarrow \mu \gamma)$ and ${\rm
BR}(\tau\rightarrow e \gamma)$ by using the reconstructed Higgs
triplet Yukawa couplings in the supersymmetric left-right model.

The remaining part of this paper is organized as follows. In
section II, we calculate the one-loop RGEs for the effective
neutrino mass operators. Section III is devoted to studying  the
lepton-flavor-violating processes. A summary of our main results
is given in section IV. Some useful formulas are listed in
appendices A and B.

\section{renormalization group equations of the effective neutrino mass operators}
We assume that the gauge and discrete left-right symmetries are
both broken by the vev of $\Delta^c$ at the high energy scale in
our model. As a result the right-handed neutrinos and Higgs
triplets are much heavier than other particles. Integrating out
the right-handed neutrinos in the leading-order approximation, one
obtains the effective neutrino mass operators, which are contained
in the $F$-term of the superpotential,
\begin{eqnarray}
{\cal W}_\kappa^{}&=&-{1\over 4}(\kappa_{\rm
1}^{})_{gf}^{}l^g_c\varepsilon^{ce} (\Phi_u^{})_{e1}^{}
l^g_a\varepsilon^{ab} (\Phi_u^{})_{b1}^{}-{1\over 4}(\kappa_{\rm
2}^{})_{gf}^{}l^f_c\varepsilon^{ce} (\Phi_d^{})_{e1}^{}
l^f_a\varepsilon^{ab} (\Phi_d^{})_{b1}^{}\nonumber\\&&-{1\over
4}(\kappa_{\rm 3}^{})_{gf}^{}l^g_c\varepsilon^{ce}
(\Phi_u^{})_{e1}^{} l^f_a\varepsilon^{ab} (\Phi_d^{})_{b1}^{}+{\rm
h.c.},
\end{eqnarray}
where
\begin{eqnarray}
\kappa_{\rm 1}^{}&=&2\left[(Y_l^u)^T (v_R^{} {\cal F})^{-1}
Y_l^u\right],\nonumber\\
\kappa_{\rm 2}^{}&=&2\left[(Y_l^d)^T (v_R^{} {\cal F})^{-1}
Y_l^d\right],\nonumber\\
\kappa_{\rm 3}^{}&=&\left[(Y_l^u)^T (v_R^{} {\cal F})^{-1}
Y_l^d+(Y_l^d)^T (v_R^{} {\cal F})^{-1} Y_l^u\right].
\end{eqnarray}
Due to the non-renormalization theorem\cite{non-renormal}, the
RGEs for operators of the superpotential are governed by the wave
function renormalization for the superfields. At the one-loop
level the wave-function renormalizaton constants $Z$ are obtained
with the dimensional regularization via the dimensional
reduction\cite{reduce}:
\begin{eqnarray}
-(4\pi)^2\delta Z_{\Phi_u^{}}^{}&=&6{\rm Tr}\left[(Y_q^u)^\dagger
Y_q^u\right]+2{\rm Tr}\left[(Y_l^u)^\dagger Y_l^u\right]-{3\over
5}g_1^2-3g_2^2\ ,\nonumber\\
-(4\pi)^2\delta Z_{\Phi_d^{}}^{}&=&6{\rm Tr}\left[(Y_q^d)^\dagger
Y_q^d\right]+2{\rm Tr}\left[(Y_l^d)^\dagger Y_l^d\right]-{3\over
5}g_1^2-3g_2^2\ ,\nonumber\\
-(4\pi)^2\delta Z_{l \ \ }^{}&=&2\left[(Y_l^u)^\dagger Y_l^{u}+
 \ (Y_l^d)^\dagger Y_l^{d}+ \ {\cal F}{\cal F^\dagger}\right]
 -{3\over 5}g_1^2-3g_2^2\ ,\nonumber\\
-(4\pi)^2\delta Z_{\Delta \ }^{}&=&4{\rm Tr}\left[{\cal F}{\cal
F^\dagger}\right]-{12\over 5}g_1^2-8g_2^2\ .
\end{eqnarray}
Using the counterterms calculated above and the technique
described in Ref. \cite{technique}, we obtain the
$\beta$-functions ($\beta_X^{}\equiv\mu{d\over d\mu}X$) of the
effective mass operators $\kappa_i^{}$ $(i=1, 2, 3)$ and the Higgs
triplet Yukawa coupling ${\cal F}$ :
\begin{eqnarray}
16\pi^2\beta_{\kappa_1^{}}^{}&=&{\cal R}^T\cdot\kappa_1^{}
+\kappa_1^{}\cdot{\cal R}+\left\{6{\rm Tr}\left[(Y_q^u)^\dagger
Y_q^u\right]-{6\over
5}g_1^2-6g_2^2\right\}\kappa_1^{}\ ,\nonumber\\
16\pi^2\beta_{\kappa_2^{}}^{}&=&{\cal R}^T\cdot\kappa_2^{}
+\kappa_2^{}\cdot{\cal R}+\left\{6{\rm Tr}\left[(Y_q^d)^\dagger
Y_q^d\right]-{6\over
5}g_1^2-6g_2^2\right\}\kappa_2^{}\ ,\nonumber\\
16\pi^2\beta_{\kappa_3^{}}^{}&=&{\cal R}^T\cdot\kappa_3^{}
+\kappa_3^{}\cdot{\cal R}+\left\{3{\rm Tr}\left[(Y_q^u)^\dagger
Y_q^u\right]+3{\rm Tr}\left[(Y_q^d)^\dagger Y_q^d\right]-{6\over
5}g_1^2-6g_2^2\right\}\kappa_3^{}\ ,\nonumber\\
16\pi^2\beta_{\cal F}^{}\ &=&{\cal R}^T\cdot{\cal F} +\ {\cal
F}\cdot{\cal R} +\left\{2{\rm Tr}\left[{\cal F}{\cal
F^\dagger}\right]-{9\over 5}g_1^2-7g_2^2\right\}{\cal F},
\end{eqnarray}
where
\begin{eqnarray}
{\cal R}\equiv(Y_l^u)^\dagger Y_l^{u}+ (Y_l^d)^\dagger Y_l^{d}+
{\cal F}{\cal F^\dagger}.
\end{eqnarray}
Some comments are in order.
\begin{itemize}

\item     In calculating the $\beta$-functions, we have assumed
$M_\Delta^{}$ (the mass of the Higgs triplet) to be lighter than
$M_1$ which is the mass of the lightest right-handed neutrinos.
Actually this assumption is not necessary.  One may integrate out
$\nu_R$ and $\Delta$ each at its own mass scale and redefining
iteratively the effective operator, which is more reasonable.
Below $m_\Delta^{}$, the $\beta$-functions of the effective mass
operators, which come from integrating out the Higgs triplet, are
similar to $\kappa_i^{}$'s.

\item     Given the vacuum expectation values of the Higgs
bi-doublets and triplets in Eq. (2), only $\kappa_1^{}$ gives rise
to masses of the light left-handed neutrinos after spontaneous
electro-weak symmetry breaking. We just need to calculate the
$\beta$-function of $\kappa_1^{}$ when considering the
renormalization group effects of neutrino mass operators. Besides,
all operators in ${\cal W}_\kappa^{}$ contribute to the
lepton-flavor-violating processes. However, such processes are
strongly suppressed by heavy masses of the right-handed neutrinos.

\item     Below the lightest seesaw scale, the $\beta$-function of
the effective  neutrino mass operator prossess the same as that of
the Type-I seesaw model in the MSSM, only up to a replacement
$Y_l^\dagger Y_l^{}$ $\longrightarrow$ $(Y_l^u)^\dagger Y_l^{u}+
(Y_l^d)^\dagger Y_l^{d}$.
\end{itemize}
Due to the renormalization group (RG) evolution effects between
the $M_\Delta^{}$ and $M_1^{}$ scales, the seesaw formula in Eq.
(3) is modified, where two ${\cal F}$'s in Type-I and Type-II
terms are not equal anymore. As a result the duality property is
slightly broken when considering the RG evolution effects of
${\cal F}$ and the effective neutrino mass operator.

\section{Lepton Flavor Violation in the Supersymmetric Left-Right Model}

In this section, we first give the analytical formulas to be used
for the calculation of the lepton-flavor-violating processes and
then list our numerical results.

\subsection{Analytical formulas}

Working in the basis where the sleptons are in weak eigenstates
together with the charginos (neutralinos) in their mass
eigenstates, we write down the interaction Lagrangian of
lepton-slepton-chargino in the following form:
\begin{eqnarray}
-{\cal
L}_{int}^{}&=&+{\tilde\nu}^\dagger_{Li}\overline{{\tilde\chi}^-_A}(C^{A(i)}_{LR}
P_R^{}+C_{LL}^{A(i)}P_L)l_i^{}+{\tilde\nu}^\dagger_{Ri}\overline{{\tilde\chi
}^-_A}(C^{A(i)}_{RR} P_R^{}+C_{RL}^{A(i)}P_L)l_i^{}\nonumber
\\&&+{\tilde e}^\dagger_{Li}\overline{\tilde\chi^0_A}(N^{A(i)}_{LR}
P_R^{}+N_{LL}^{A(i)}P_L)l_i^{}+{\tilde
e}^\dagger_{Ri}\overline{\tilde\chi^0_A}(N^{A(i)}_{RR}
P_R^{}+N_{RL}^{A(i)}P_L)l_i^{}\\&&+{\rm h.c.},\nonumber
\end{eqnarray}
where the coefficients are:
\begin{eqnarray}
C_{LL}^{A(i)}&=&g_L^{}({\cal O}_R^{})_{A1}^{},\nonumber\\
C_{LR}^{A(i)}&=&-{g_L^{}m_{ei}^{}\over \sqrt{2}m_W^{} \cos
\beta}({\cal O}_L^{})_{A3}^{}+{g_L^{}m^D_{\nu i}\over
\sqrt{2}m_W^{} \sin \beta}({\cal O}_L^{})_{A4}^{},\nonumber\\
C_{RR}^{A(i)}&=&g_R^{}({{\cal O}_L})_{A2}^{},\nonumber\\
C_{LR}^{A(i)}&=&-{g_L^{}m_{ei}^{}\over \sqrt{2}m_W^{} \cos
\beta}({\cal O}_R^{})_{A3}^{}+{g_L^{}m^D_{\nu i}\over
\sqrt{2}m_W^{} \sin \beta}({\cal O}_R^{})_{A4}^{},\nonumber\\
N_{LL}^{A(i)}&=&{g_L\over \sqrt{2}}\left[-({{\cal
O}_N^{}})_{A2}^{}-({{\cal O}_N^{}})_{A1}\tan\theta_W^{}\right],\nonumber\\
N_{LR}^{A(i)}&=&
{g_L^{}m_{ei}^{}\over\sqrt{2}m_W^{}\cos\beta}\left[({{\cal
O}_N^{}})_{A3}-({{\cal O}_N^{}})_{A4}\right]+{g_L^{}m^D_{\nu
i}\over \sqrt{2}m_W^{} \sin \beta}\left[({{\cal
O}_N^{}})_{A6}-({{\cal O}_N^{}})_{A5}\right],\nonumber\\
N_{RL}^{A(i)}&=&N_{LR}^{A(i)},\nonumber\\
N_{RR}^{A(i)}&=&{g_R\over\sqrt{2}}\left[-({{\cal
O}_N^{}})_{A7}-({{\cal O}_N^{}})_{A1}\tan \theta_W^{}\right].
\end{eqnarray}
Here ${\cal O}_L^{}$, ${\cal O}_R^{}$ and ${\cal O}_N^{}$ are real
orthogonal matrices that diagonalize chargino and neutralino mass
matrices respectively. Their explicit forms are listed in appendix
A. $\tan \beta\equiv \kappa_u^{}/\kappa_d^{}$ is defined.

Let us discuss the branching ratios  of the
lepton-flavor-violating processes in the supersymmetric left-right
model. The radiative decays $l_i^{}\rightarrow l_j^{}+\gamma$ are
induced by the effective operator\cite{xianzhi}:
\begin{eqnarray}
e\overline{l_j^{}} \left(iD_L^\gamma P_L^{}+iD_R^\gamma
P_R^{}\right)\sigma^{\mu\nu}l_i^{} F_{\mu\nu}^{}+{\rm h.c.}\ ,
\end{eqnarray}
where $e$ and $F_{\rho\sigma}^{}$ are the charge and the
electromagnetic field strength, respectively. These operators are
chirality-flipping (dipole) and come from $SU(2)_{\rm L}^{}\times
U(1)_{\rm Y}^{}$-invariant operators with at least one Higgs
field.

In the ``mass insertion" method and leading-log approximations,
the coefficients $D^\gamma_{L,R}$ can be
calculated\cite{theotrical} and we write down the explicit
expression  in appendix A. The branching ratio of $l_i\rightarrow
l_j+\gamma$ decay due to the new contributions is given by:
\begin{eqnarray}
{\rm BR}(l_i^{}\rightarrow l_j^{}\gamma)={48\pi^3\alpha\over
m_{l_i}^2G_F^2}\left(\mid D_L^\gamma\mid^2+\mid
D_R^\gamma\mid^2\right){\rm BR}(l_i^{}\longrightarrow
l_j^{}\bar{\nu}_j^{}\nu_i^{}),
\end{eqnarray}
where $\alpha=e^2/(4\pi)$, $G_F^{}$ is the Fermi constant, ${\rm
BR}(\tau\rightarrow\mu\nu_\tau^{}\overline{\nu}_\mu^{})\approx17\%$
and ${\rm BR}(\tau\rightarrow
e\nu_\tau^{}\overline{\nu}_e^{})\approx18\%$\cite{pdg}.

In the minimal SUGRA scenario, at the gravitational scale the
supersymmetry breaking masses for sleptons, squarks and the Higgs
bosons are universal, and the SUSY breaking parameters associated
with the supersymmetric Yukawa couplings or masses are
proportional to the Yukawa coupling constants or masses. Then, the
SUSY breaking parameters are given as:
\begin{eqnarray}
&&(m_L^2)_{ij}=(m_R^2)_{ij}=(m_\nu^2)_{ij}=\delta_{ij}m_0^2,\nonumber\\
&&m_{\tilde{\Phi}_1}^2=m_{\tilde{\Phi}_2}^2=m_0^2,\nonumber\\
&&(A_{l}^{u,d})^{ij}=(Y_{l}^{u,d})^{ij}a_0^{},
A_{{\cal F}}^{ij}={\cal F}^{ij}a_0^{},\nonumber\\
&&B_\nu^{ij}=M_{\nu_i\nu_j}b_0, B_\Phi=\mu b_0.
\end{eqnarray}
Flavor violation in the slepton sector arises from radiative
corrections induced by the flavor-violating couplings of heavy
states populating the theory between the Planck scale and the
electroweak scale. Integrating the one-loop renormalization group
equations \cite{RGE} for the soft breaking masses $m_L^2$, $m_R^2$
and  trilinear $A_l^{u, d}$ in the lowest-order approximation, one
obtains the off-diagonal term for  $m_L^2$, $m_R^2$ and $A_l^{u,
d}$:
\begin{eqnarray}
(m_L^2)_{ij}\simeq(m_R^2)_{ij}\simeq-{3m_0^2+a_0^2\over
4\pi^2}{\cal R}_{ij}^{},\hspace{1 cm}A_l^{u, d} \simeq-{3\over
4\pi^2}Y_l^{u, d} a_0^{}{\cal R}_{ij}^{},
\end{eqnarray}
where
\begin{eqnarray}
{\cal R}_{ij}^{}=\left[
Y_l^u(Y_l^u)^\dagger+Y_l^d(Y_l^d)^\dagger\right]_{ij}^{} {\log}
\left({M_P\over M_R}\right)+3({\cal F}{\cal
F}^\dagger)_{ij}^{}{\log}\left({M_P\over
M_\Delta}\right).\nonumber
\end{eqnarray}
These off-diagonal terms generate new contributions in the
amplitudes of lepton-flavor-violating processes\cite{lfv} such as
$\tau\rightarrow \mu \gamma$ and $\tau\rightarrow e \gamma$ .

\subsection{Numerical results}

The lepton flavor mixing matrix ($U_{\rm MNS}^{}$) comes from the
mismatch between the diagonalizations of the neutrino mass matrix
and the charged lepton mass matrix. The tri-bimaximal mixing
pattern\cite{TB} is strongly favored by the solar and atmospheric
neutrino oscillation measurements:
\begin{eqnarray}
U_{\rm MNS}^{}=\left(\matrix{{2\over\sqrt{6}}&{1\over \sqrt{3}}&0\cr
-{1\over \sqrt{6}}&{1\over \sqrt{3}}&{1\over \sqrt{2}}\cr -{1\over
\sqrt{6}}&{1\over \sqrt{3}}&-{1\over \sqrt{2}}}\right).
\end{eqnarray}
A global analysis of current experimental data yields the values
for the solar mass splitting  $\Delta m_{12}^2= (8.0\pm0.3) \times
10^{-5} {\rm eV^2}$ and the atmospheric mass splitting $|
m_{23}^2|=(2.5\pm0.2)\times 10^{-3} {\rm eV^2}$\cite{Vissani}. We
assume that three light left-handed Majorana neutrinos are in
normal mass hierarchy (i.e., $m_1^{}< m_2^{}<m_3^{}$), so that
$m_3^{}\simeq \sqrt{\mid \Delta m_{23}^2\mid}\simeq 0.05 {\rm eV}$
and $m_2^{}\simeq\sqrt{\Delta m_{12}^2} \simeq 0.009 {\rm eV}$. We
also take $m_1^{}\simeq 0.001 {\rm eV}$, $v_L^{}\simeq 0.05 {\rm
eV}$ and $v_L^{} v_R^{}/v_u^2 \simeq 1$, which are natural
values\cite{gamma}.

We assume that at the GUT scale the theory is given by the
supersymmetric  $SO(10)$ model which contains two 10-dimensional
and a pair of $126\oplus\overline{126}$ representation Higgs
bosons. Then the most general Yukawa couplings lead to the
following mass relation for the fermions: $M_u^{}=M_D^{}$. We
neglect the CKM relations between the up- and down-type quarks in
our numerical calculations, assuming that the up-type and
down-type quark mass matrices are both diagonal. The Dirac
neutrino mass matrix turns out to be $M_D^{}= {\rm diag}(m_u^{},
m_c^{}, m_t^{})$.

Using these choices and the technique described in \cite{first},
one obtains eight different solutions for the triplet Yukawa
coupling ${\cal F}$ through  the left-right seesaw formula in Eq.
(3):
\begin{eqnarray}
{\cal F}_1^{}&\simeq&\left(\matrix{-0.00169&-0.00349 &0.00015\cr
-0.00349&0.51022&-0.51309\cr 0.00015&-0.51309&0.69097}\right),
\hspace{0.4cm}
{\cal F'}_1^{}\simeq\left(\matrix{0.06236&0.06316 &0.05952\cr
0.06316&0.04995&0.07326\cr 0.05952&0.07326&-0.13080}\right),
\nonumber\\
{\cal F}_2^{}&\simeq&\left(\matrix{0.06235&0.06316 &0.05950\cr
0.06316&0.04996&0.07515\cr 0.05950&0.07515&0.21616}\right),
\hspace{0.9cm}
{\cal F'}_2^{}\simeq\left(\matrix{-0.00169&-0.00349 &0.00016\cr
-0.00349&0.51021&-0.51498\cr 0.00016&-0.51498&0.34400}\right),
\nonumber\\
\nonumber\\
{\cal F }_3^{}&\simeq&\left(\matrix{-4\cdot 10^{-10}&
4\cdot10^{-8} &6\cdot 10^{-6}\cr 4\cdot
10^{-8}&-7\cdot10^{-6}&-9\cdot10^{-4}\cr
6\cdot10^{-6}&-9\cdot10^{-4}&-0.1736}\right), \hspace{0.2cm}
{\cal F'}_3^{}\simeq\left(\matrix{0.06067&0.05967 &0.05967\cr
0.05967&0.56017&-0.43888\cr 0.05966&-0.43888&0.73374}\right),
\nonumber\\
{\cal F }_4^{}&\simeq&\left(\matrix{5\cdot 10^{-11}&
-3\cdot10^{-8} &-6\cdot 10^{-6}\cr -3\cdot
10^{-8}&3\cdot10^{-6}&9\cdot10^{-4}\cr
-6\cdot10^{-6}&9\cdot10^{-4}&0.17342}\right), \hspace{0.3cm}
{\cal F'}_4^{}\simeq\left(\matrix{0.06067&0.05966 &0.05967\cr
0.05966&0.56016&-0.44078\cr 0.05967&-0.44078&0.38678}\right).
\nonumber\\
\end{eqnarray}
It is easy to check that  the duality relation (${\cal F}_i^{}
$+${\cal F'}_i^{}$= $m_\nu^{}/v_L^{}$) is satisfied very
accurately for the solutions given above.

Now, we present our numerical results of ${\rm BR}
(\tau\rightarrow \mu, e+ \gamma)$ in the parameter space given
above. The experimental upper limits on those branching ratios
are: ${\rm BR}(\tau \rightarrow \mu+\gamma)<6.8\times 10^{-8}$ and
$ {\rm BR}(\tau \rightarrow e+\gamma)<1.1\times 10^{-7}$ at $90
\%$ C.L.\cite{new data} and the sensitivities of a few planned
experiments\cite{future} may reach ${\rm BR}(\tau\rightarrow
e+\gamma)\sim {\cal O} (10^{-8})$ and ${\rm BR}(\tau\rightarrow
\mu+\gamma)\sim {\cal O} (10^{-8})$. FIG. 1 and FIG. 2 show the
${\rm BR}(\tau\rightarrow[ \mu, e ] +\gamma)$ changing with
$m_0^{}$. We find that the experimentally allowed ranges of ${\rm
BR}(\tau\rightarrow [\mu, e ]+\gamma)$ can be reproduced from all
of these eight different triplet Yukawa couplings in the chosen
parameter space. Besides, curves corresponding to ${\cal F}_3^{}$
and ${\cal F}_4^{}$ are lapped over with each other because there
is little difference in their numerical expression. Although eight
different Higgs triplet Yukawa couplings result in the same
neutrino mass matrix through the Type-II seesaw formula, their
effects on lepton-flavor-violating processes are very different.
As a result, we may check the stability of the Type-II seesaw
formula by measuring the branching ratios of the
lepton-flavor-violating $\tau$ decays accurately in the future
experiments.

\section{summary}
In addition to the right-handed neutrinos, the Higgs triplet is
another source of the neutrino mass generation in the Type-II
seesaw model, so the evolution of the neutrino mass matrix is a
little different from that in the Type-I seesaw model. Besides,
the duality property for the Type-II seesaw formula indicates that
there exist eight possible Higgs triplet Yukawa couplings ${\cal
F}$ which , for a given $M_D^{}$, result in exactly the same mass
matrix of light neutrinos. In this article, we have calculated the
RGEs for the evolutions of the Type-II seesaw neutrino mass
matrices from the seesaw scale to the electro-weak scale in the
supersymmetric left-right model. Instead of giving numerical
analysis, we have discussed the stability of the Type-II seesaw
model. On the other hand, the Higgs triplet Yukawa coupling is an
important source for the lepton-flavor-violating $\tau$ decays. We
have calculated these eight Yukawa couplings through the Type-II
seesaw formula and applied them to evaluating the branching ratios
of lepton-flavor-violating $\tau$ decays. We find that their
contributions to the branching ratios are different and the
stability of the Type-II seesaw can be checked by measuring rare
$\tau$ decay accurately.

In conclusion, the supersymmetric left-right model supplies an
interesting platform for the neutrino sector, which could be
tested in the future LHC and ILC experiments.

\begin{acknowledgments}
The author is indebted to Professor Zhi-zhong Xing for reading the
manuscript with great care and patience, and also for his valuable
comments and numerous corrections. He is also grateful to S. Zhou
and H. Zhang for useful discussions. This work was supported in
part by the National Nature Science Foundation of China.
\end{acknowledgments}
\pagebreak

\appendix {\section{}

In this appendix, we consider chargino mixing and neutralino
mixing in the supersymmetric left-right model. We first write down
the $\lambda-\phi-A$ terms of the Lagrangian, which involve the
soft supersymmetry-breaking terms and the scalar
potential\cite{lrsusy1,ncmixing}.
\begin{eqnarray}
\ell_{GH}&=&+i\sqrt{2}{\rm
Tr}[(\sigma\cdot\Delta_L)^\dagger(g_L\sigma\cdot\lambda_L+2g_v
\lambda_v)\sigma\cdot\tilde{\Delta}_L]+h.c.\nonumber \\
&&+i\sqrt{2}{\rm
Tr}[(\sigma\cdot\Delta_R)^\dagger(g_R\sigma\cdot\lambda_R+2g_v
\lambda_v)\sigma\cdot\tilde{\Delta}_R]+h.c.\nonumber \\
&&+{i\over \sqrt{2}}{\rm
Tr}[\Phi_u^\dagger(g_L\sigma\cdot\lambda_L+g_R\sigma\cdot\lambda_R)\tilde{\Phi}_u]+h.c.
\nonumber \\&&+{i\over \sqrt{2}}{\rm
Tr}[\Phi_d^\dagger(g_L\sigma\cdot\lambda_L+g_R\sigma\cdot\lambda_R)\tilde{\Phi}_d]+h.c.
\nonumber \\&&+{\rm
Tr}[\mu_2(\sigma\cdot\tilde{\Delta}_L)(\sigma\cdot\tilde{\delta}_L)]+{\rm
Tr}[\mu_3(\sigma\cdot\tilde{\Delta}_R)(\sigma\cdot\tilde{\delta}_R)]+h.c.\nonumber
\\&&+m_L(\lambda_L^\alpha\lambda_L^\alpha+\bar{\lambda}_L^\alpha\bar{\lambda}_L^\alpha
)+m_R(\lambda_R^\alpha\lambda_R^\alpha+\bar{\lambda}_R^\alpha\bar{\lambda}_R^\alpha
)+m_v(\lambda_v\lambda_v+\bar{\lambda}_v\bar{\lambda}_v )\nonumber
\\&&+{\rm
Tr}[\mu_1(\sigma_1\tilde{\Phi}_u\sigma_1)^T\tilde{\Phi}_d].
\end{eqnarray}
Substituting the vacuum expectation values of the Higgs fields
from Eq. (2) into Eq. (17), Keeping only the terms involving
charged fields, we get
\begin{eqnarray}
\ell_C&=&\left\{i\lambda_R^-(\sqrt{2}g_R^{}v_R^{}\tilde{\Delta}^
\dagger_R+g_Rk_d\tilde{\phi}_d^\dagger)+i\lambda_L^-(\sqrt{2}g_L^{}v_L^{}\tilde{\Delta}^
\dagger_L+g_Lk_d\tilde{\phi}_d^\dagger)\right.\nonumber\\
&&+i\lambda^\dagger_Rg_Rk_u\tilde{\phi}_u^-+i\lambda_L^\dagger g_L
k_u\tilde{\phi}_u^-+4m_L^{}\lambda_L^\dagger\lambda_L^-+4m_R^{}\lambda_R^\dagger\lambda_R^-+\mu_1\tilde{\phi}_u^\dagger
\tilde{\phi}_d^-\nonumber\\&&\left.+\mu_1\tilde{\phi}_u^-\tilde{\phi}_d^++\mu_2\tilde{\Delta}_L^+\tilde{\delta}_L^-
+\mu_3\tilde{\Delta}_R^-\tilde{\delta}_R^- \right\}+{\rm h.c.} .
\end{eqnarray}
We consider the chargino mass matrix $M_C$, which is a $6 \times
6$ matrix appearing in the chargino mass terms.
\begin{eqnarray}
\ell_C&=&-{1\over 2}(\psi^{+T},\psi^{-T})\left(\matrix{0&M_C^T\cr
M_C&0\cr}\right)\left(\matrix{\psi^{+}\cr\psi^{-}\cr}\right)+{\rm
h.c.},
\end{eqnarray}
In this model, $\psi$ is defined to stand for the following
fields:
\begin{eqnarray}
\psi^+\equiv&&\left(-i\lambda_L^+, -i\lambda_R^+,
\tilde{\phi}_u^+,
\tilde{\phi}_d^+, \tilde{\Delta}_L^+, \tilde{\Delta}_R^+\right)^T,\\
\psi^-\equiv&&\left(-i\lambda_L^-, -i\lambda_R^-,
\tilde{\phi}_u^-, \tilde{\phi}_d^-, \tilde{\delta}_L^-,
\tilde{\delta}_R^-\right)^T.
\end{eqnarray}
Comparing  Eq. (19) with Eq. (17), we write down the explicit
expression of $M_C^{}$:
\begin{eqnarray}
M_C=\left(\matrix{4m_L&0&0&g_Lk_d&\sqrt{2}g_Lv_L&0\cr0&4m_R&0&g_Rk_d&0&\sqrt{2}g_Rv_R\cr
g_Lk_u&g_R k_u&0&\mu_1&0&0\cr0&0&\mu_1&0&0&0\cr0&0&0&0&\mu_2&0\cr
0&0&0&0&0&\mu_3\cr}\right).
\end{eqnarray}
By defining  $\chi_i^-={\cal O}_R^*\psi^-$,$\chi^+={\cal
O}_L^{*}\psi^+$, we can diagonalize $M_C$ by $6\times 6$
orthogonal matrices ${\cal O}_R^{}$ and ${\cal O}_L^{}$ according
to ${\cal O}_R^{}M_C^{}{\cal O}_L^{T}=M_C^D$, where $M_C^D$ is a
diagonal matrix. It is tedious to write down the analytical
expressions of ${\cal O}_L^{}$ and ${\cal O}_R^{}$. Hence we only
list their numerical expressions:
\begin{eqnarray}
{\cal
O}_R^{}&\approx&\left(\matrix{0&-0.999&-0.002&0&0&-0.009\cr0.996&
0&0.090&0&0&-0.002\cr0.001&0&0&-1&0&0\cr-0.075&0.005&
0.817&0&0&-0.572\cr-0.050&-0.008&0.570&0&0&0.820\cr0&0&0&0&1&0
}\right)\ , \nonumber\\
{\cal O}_L^{}&\approx&\left(\matrix{0&0.196&0&0.001&0&0.981
\cr0.998&0.009&0&0.062&0.001&-0.002\cr-0.001&0&-0.371&0
&0.929&0\cr-0.034&-0.739&0&0.656&0&0.147\cr0.053&-0.644
&0&-0.752&0&0.129\cr0&0&0.929&0&0.371&0}\right).
\end{eqnarray}
Here we choose $M_L^{}=1$ TeV, $M_R^{}=20$ TeV,
$\mu_1=\mu_2=\mu_3=200$ GeV, $\tan \beta=1.5$, $v_L^{}=0.05$ eV
and $v_R^{}=10^{10}$ GeV in our calculation.

In  order to obtain the neutralino part of the Lagrangian, we
replace the vevs of the Higgs bosons into Eq. (17), keeping only
the neutral terms:
\begin{eqnarray}
\ell_N&=&\left\{-i\sqrt{2}(\lambda_L^0
g_L-2\lambda_v^0g_V)v_L\tilde{\Delta}_L^0-i\sqrt{2}(\lambda_R^0
g_R-2\lambda_v^0g_V)v_R\tilde{\Delta}_R^0\right.\nonumber\\
&&+i{1\over
\sqrt{2}}(\lambda_R^0g_R-\lambda_L^0g_L)\kappa_u\tilde{\phi}^0_{1u}
-i{1\over
\sqrt{2}}(\lambda_R^0g_R-\lambda_L^0g_L)\kappa_d\tilde{\phi}^0_{2d}
\nonumber\\&&+m_L(\lambda_L^0\lambda_L^0+\bar{\lambda}_L^0\bar\lambda_L^0)
+m_R(\lambda_R^0\lambda_R^0+\bar{\lambda}_R^0\bar\lambda_R^0)+m_V(\lambda_V^0
\lambda_V^0+\bar{\lambda}_V^0\bar\lambda_V^0)\nonumber\\
&&\left.+\mu_1(\tilde{\phi}_{1u}^0\tilde{\phi}_{2d}^0+\tilde{\phi}_{2u}^0\tilde{\phi}_{1d}^0)
\right\}+{\rm h.c.}.
\end{eqnarray}

The neutralino particles are produced in two stages of symmetry
breaking\cite{neutralino}. The first stage,  the vev $v_R$
  is responsible for giving masses to the heavy neutralinos. The
second stage, the vevs $\kappa_u$ and $\kappa_d$  are responsible
for giving masses to the light neutralinos. The amount of mixing
between heavy and light neutralinos is small, so one can calculate
the neutralino mass eigenstates for both stages as independent
cases.

We define $\xi_N^{}$:
\begin{eqnarray}
\xi_N^{}\equiv(-i\lambda_L^0,-i\lambda_R^0,\tilde{\phi}_{1u}^0
,\tilde{\phi}_{2u}^{0},\tilde{\phi}_{1d}^0
,\tilde{\phi}_{2d}^{0}),
\end{eqnarray}
Then the relevant part in  Eq. (24) may be written as:
\begin{eqnarray}
\ell_N=-{1\over 2}\xi_NM_N\xi_N^T+{\rm h.c.},
\end{eqnarray}
where
\begin{eqnarray}
M_N^{}=
\left(\matrix{m_L&0&{-1\over\sqrt{2}}g_L\kappa_u&0&0&{1\over\sqrt{2}}g_L\kappa_d
&\cr0&m_R&{1\over \sqrt{2}}g_R\kappa_u&0&0&{-1\over
\sqrt{2}}g_R\kappa_d\cr
{-1\over\sqrt{2}}g_L\kappa_u&{1\over\sqrt{2}}g_R\kappa_u&0&0&0&-\mu_1&\cr
0&0&0&0&-\mu_1&0&\cr0&0&0&-\mu_1&0&0\cr
{1\over\sqrt{2}}g_l\kappa_d&{-1\over\sqrt{2}}g_R\kappa_d&-\mu_1&0&0&0\cr
}\right).
\end{eqnarray}
$M_N^{}$ is diagonalized by a real orthogonal matrix ${\cal
O}_N^{}$ with ${\cal O}_N^{}M_N^{}{\cal O}_N^{T}=M_N^D$. We write
down the numerical expression for ${\cal O}_N^{}$:
\begin{eqnarray}
{\cal O}_N^{}=\left(\matrix{0&0.999&0.005&0&0&0.004\cr-0.995&0.001
&-0.088&0&0&-0.061\cr-0.106&-0.006&0.707&0&0&0.700\cr0&0&0&-0.707&
0.707&0\cr0&0&0&-0.707&-0.707&0\cr0.018&0.001&-0.702&0&0&0.711}\right).
\end{eqnarray}
Here we choose $M_L^{}=1$ TeV, $M_R^{}=20$ TeV, $\mu_1=200$ GeV
and $\tan \beta=1.5$ in our calculation.

\section{}
 In this appendix, we write down the formula of $D^\gamma_{L,R}$
\footnote{we do not consider the contributions of the double
charged chargino mediated diagrams, since their contributions are
very small.},
which are a little different from the formula given in Ref.
\cite{theotrical}:
\begin{eqnarray}
D_L^\gamma&=&-{1\over2(4\pi)^2}M_{\tilde\chi^0}N_{RR}^{A(i)}N_{LL}^{A(j)}A_{e}^{ii}(\bar{m}_{\tilde
e}^2)_{ij}\left({1\over m_{\tilde e_{Ri}}^2-m_{\tilde
e_{Li}}^2}{1\over m_{\tilde e_{Ri}}^2-m_{\tilde
e_{Lj}}^2}{g_n(M_{\tilde \chi^0}^2/\bar{m}_{\tilde e_{Ri}}^2)\over
m_{\tilde e_{Ri}}^2}\right. \nonumber \\&&\left.+{1\over m_{\tilde
e_{Li}}^2-m_{\tilde e_{Ri}}^2}{1\over m_{\tilde
e_{Li}}^2-m_{\tilde e_{Lj}}^2}{g_n(M_{\tilde
\chi^0}^2/\bar{m}_{\tilde e_{Li}}^2)\over m_{\tilde e_{Li}}^2}+
{1\over m_{\tilde e_{Lj}}^2-m_{\tilde e_{Ri}}^2}{1\over m_{\tilde
e_{Lj}}^2-m_{\tilde e_{Li}}^2}{g_n(M_{\tilde
\chi^0}^2/\bar{m}_{\tilde e_{Lj}}^2)\over m_{\tilde
e_{Lj}}^2}\right) \nonumber\\ &&-{1\over
6(4\pi)^2}m_{e_i}N_{LL}^{A(i)}N_{LL}^{A(j)} {(\bar{m}_{\tilde
e}^2)_{ij}\over \bar{m}_{\tilde e_i}^2-\bar{m}_ {\tilde
e_j}^2}\left({f_n(M_{\tilde \chi^0}^2/\bar{m}_{\tilde e_i}^2)\over
\bar{m}_{\tilde e_i}^2}- {f_n(M_{\tilde \chi^0}^2/\bar{m}_{\tilde
e_j}^2)\over \bar{m}_{\tilde e_j}^2} \right)
\nonumber\\&&-{1\over2(4\pi)^2}M_{\tilde\chi^0}N_{LR}^{A(i)}N_{LL}^{A(j)}
{(\bar{m}_{\tilde e}^2)_{ij}\over \bar{m}_{\tilde e_i}^2-\bar{m}_
{\tilde e_j}^2}\left({g_n(M_{\tilde \chi^0}^2/\bar{m}_{\tilde
e_i}^2)\over \bar{m}_{\tilde e_i}^2}- {g_n(M_{\tilde
\chi^0}^2/\bar{m}_{\tilde e_j}^2)\over \bar{m}_{\tilde e_j}^2}
\right)\nonumber\\&&+{1\over
(4\pi)^2}M_{\chi^-_{A}}C_{RR}^{A(i)}C_{LL}^{A(j)}A_\nu^{ii}(\tilde{m}_L^2)_{ij}
\left({1\over m_{\nu_{Ri}}^2-m_{\nu_{Li}}^2}{1\over
m_{\nu_{Ri}}^2-m_{\nu_{Lj}}^2}{g_c(M_{\tilde
\chi^-}^2/\bar{m}_{\tilde \nu_{Ri}}^2)\over m_{\nu_{Ri}}^2}
\nonumber\right.\\ &&\left.+{1\over
m_{\nu_{Li}}^2-m_{\nu_{Ri}}^2}{1\over
m_{\nu_{Li}}^2-m_{\nu_{Lj}}^2}{g_c(M_{\tilde
\chi^-}^2/\bar{m}_{\tilde \nu_{Li}}^2)\over
m_{\nu_{Li}}^2}+{1\over m_{\nu_{Lj}}^2-m_{\nu_{Ri}}^2}{1\over
m_{\nu_{Lj}}^2-m_{\nu_{Li}}^2}{g_c(M_{\tilde
\chi^-}^2/\bar{m}_{\tilde \nu_{Lj}}^2)\over m_{\nu_{Lj}}^2}\right)
\nonumber\\&&+{1\over
6(4\pi)^2}m_{e_i}C_{LL}^{A(i)}C_{LL}^{A(j)}{(\bar{m}_{\tilde
e}^2)_{ij}\over \bar{m}_{\tilde \nu_i}^2-\bar{m}_{\tilde
\nu_j}^2}\left({f_c(M_{\tilde \chi^-}^2/\bar{m}_{\tilde
\nu_i}^2)\over \bar{m}_{\tilde \nu_i}^2}- {f_c(M_{\tilde
\chi^-}^2/\bar{m}_{\tilde \nu_j}^2)\over \bar{m}_{\tilde \nu_j}^2}
\right)\nonumber\\&&+{1\over
(4\pi)^2}M_{\tilde\chi^-}C_{LR}^{A(i)*}C_{LL}^{A(j)}{(\bar{m}_{\tilde
e}^2)_{ij}\over \bar{m}_{\tilde \nu_i}^2-\bar{m}_{\tilde
\nu_j}^2}\left({g_c(M_{\tilde \chi^-}^2/\bar{m}_{\tilde
\nu_i}^2)\over \bar{m}_{\tilde \nu_i}^2}- {g_c(M_{\tilde
\chi^-}^2/\bar{m}_{\tilde \nu_j}^2)\over \bar{m}_{\tilde \nu_j}^2}
\right),
\end{eqnarray}
\begin{eqnarray}
D_R^\gamma&=&-{1\over2(4\pi)^2}M_{\tilde\chi^0}N_{LL}^{A(i)}N_{RR}^{A(j)}A_{e}^{ii}(\bar{m}_{\tilde
e}^2)_{ij}\left({1\over m_{\tilde e_{Ri}}^2-m_{\tilde
e_{Li}}^2}{1\over m_{\tilde e_{Ri}}^2-m_{\tilde
e_{Rj}}^2}{g_n(M_{\tilde \chi^0}^2/\bar{m}_{\tilde e_{Ri}}^2)\over
m_{\tilde e_{Ri}}^2}\right. \nonumber \\&&\left.+{1\over m_{\tilde
e_{Li}}^2-m_{\tilde e_{Ri}}^2}{1\over m_{\tilde
e_{Li}}^2-m_{\tilde e_{Rj}}^2}{g_n(M_{\tilde
\chi^0}^2/\bar{m}_{\tilde e_{Li}}^2)\over m_{\tilde e_{Li}}^2}+
{1\over m_{\tilde e_{Rj}}^2-m_{\tilde e_{Ri}}^2}{1\over m_{\tilde
e_{Rj}}^2-m_{\tilde e_{Li}}^2}{g_n(M_{\tilde
\chi^0}^2/\bar{m}_{\tilde e_{Rj}}^2)\over m_{\tilde
e_{Rj}}^2}\right)\nonumber\\&&-{1\over
6(4\pi)^2}m_{e_i}N_{RR}^{A(i)}N_{RR}^{A(j)} {(\bar{m}_{\tilde
e_R}^2)_{ij}\over \bar{m}_{\tilde e_{Ri}}^2-\bar{m}_ {\tilde
e_{Rj}}^2}\left({f_n(M_{\tilde \chi^0}^2/\bar{m}_{\tilde
e_{Ri}}^2)\over \bar{m}_{\tilde e_{Ri}}^2}- {f_n(M_{\tilde
\chi^0}^2/\bar{m}_{\tilde e_{Rj}}^2)\over \bar{m}_{\tilde
e_{Rj}}^2} \right)
\nonumber\\&&-{1\over2(4\pi)^2}M_{\tilde\chi^0}N_{RL}^{A(i)}N_{RR}^{A(j)}
{(\bar{m}_{\tilde e_R}^2)_{ij}\over \bar{m}_{\tilde
e_{Ri}}^2-\bar{m}_ {\tilde e_{Rj}}^2}\left({g_n(M_{\tilde
\chi^0}^2/\bar{m}_{\tilde e_{Ri}}^2)\over \bar{m}_{\tilde
e_{Ri}}^2}- {g_n(M_{\tilde \chi^0}^2/\bar{m}_{\tilde
e_{Rj}}^2)\over \bar{m}_{\tilde e_{Rj}}^2}
\right)\nonumber\\&&+{1\over
(4\pi)^2}M_{\chi^-_{A}}C_{LL}^{A(i)}C_{RR}^{A(j)}A_\nu^{ii}(\tilde{m}_R^2)_{ij}
\left({1\over m_{\nu_{Ri}}^2-m_{\nu_{Li}}^2}{1\over
m_{\nu_{Ri}}^2-m_{\nu_{Rj}}^2}{g_c(M_{\tilde
\chi^-}^2/\bar{m}_{\tilde \nu_{Ri}}^2)\over m_{\nu_{Ri}}^2}
\nonumber\right.\\ &&\left.+{1\over
m_{\nu_{Li}}^2-m_{\nu_{Ri}}^2}{1\over
m_{\nu_{Li}}^2-m_{\nu_{Rj}}^2}{g_c(M_{\tilde
\chi^-}^2/\bar{m}_{\tilde \nu_{Li}}^2)\over
m_{\nu_{Li}}^2}+{1\over m_{\nu_{Rj}}^2-m_{\nu_{Ri}}^2}{1\over
m_{\nu_{Rj}}^2-m_{\nu_{Li}}^2}{g_c(M_{\tilde
\chi^-}^2/\bar{m}_{\tilde \nu_{Rj}}^2)\over m_{\nu_{Rj}}^2}\right)
\nonumber\\&&+{1\over
6(4\pi)^2}m_{e_i}C_{RR}^{A(i)}C_{LL}^{A(j)}{(\bar{m}_{\tilde
e_R}^2)_{ij}\over \bar{m}_{\tilde \nu_{Ri}}^2-\bar{m}_{\tilde
\nu_{Rj}}^2}\left({f_c(M_{\tilde \chi^-}^2/\bar{m}_{\tilde
\nu_{Ri}}^2)\over \bar{m}_{\tilde \nu_{Ri}}^2}- {f_c(M_{\tilde
\chi^-}^2/\bar{m}_{\tilde \nu_{Rj}}^2)\over \bar{m}_{\tilde
\nu_{Rj}}^2}\right)\nonumber\\&&+{1\over
(4\pi)^2}M_{\tilde\chi^-}C_{RL}^{A(i)}C_{RR}^{A(j)}{(\bar{m}_{\tilde
e_R}^2)_{ij}\over \bar{m}_{\tilde \nu_{Ri}}^2-\bar{m}_{\tilde
\nu_{Rj}}^2}\left({g_c(M_{\tilde \chi^-}^2/\bar{m}_{\tilde
\nu_{Ri}}^2)\over \bar{m}_{\tilde \nu_{Ri}}^2}- {g_c(M_{\tilde
\chi^-}^2/\bar{m}_{\tilde \nu_{Rj}}^2)\over \bar{m}_{\tilde
\nu_{Rj}}^2} \right),
\end{eqnarray}
where the loop functions are
\begin{eqnarray}
f_n(x)&=&-{1\over 2(1-x)^4}(2+3x-6x^2+x^3+6x{\log
}x),\nonumber\\g_n(x)&=&-{1\over (1-x)^3}(1-x^2+2x{\log}x),
\nonumber\\ f_c(x)&=&-{1\over
2(1-x)^4}(2+3x-6x^2+x^3+6x{\log}x),\nonumber \\g_c(x)&=&{1\over
2(1-x)^3}(3-4x+x^2+2{\log}x).
\end{eqnarray}}

\begin{figure}
\begin{center}
\vspace{-0.5cm}
\includegraphics[]{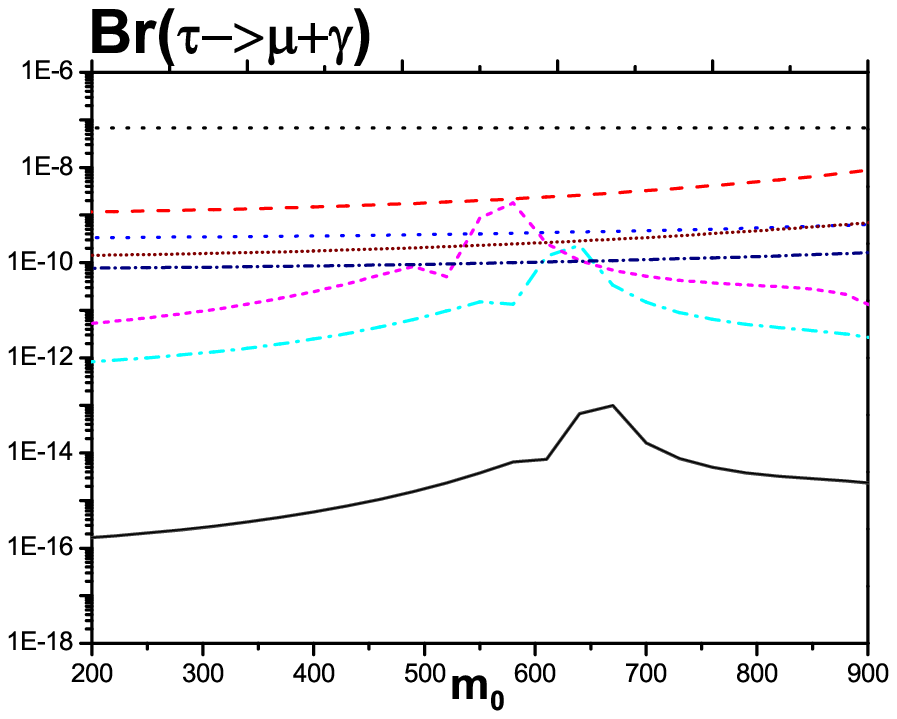} \caption{ Illustrative plot for ${\rm BR}(\tau\rightarrow\mu+\gamma)$
changing with $m^{}_{0}$.  We take $\tan\beta=1.5$, $M_L^{}=1\ {
\rm Tev}$ and $M_R^{}=20\ {\rm Tev}$ in our plot. Here the dot
line corresponds to ${\cal F}_1^{}$; the dash dot line corresponds
to ${\cal F'}_1^{}$; the short dash line corresponds to ${\cal
F}_2^{}$; the short dash dot line corresponds to ${\cal F'}_2^{}$;
the solid line corresponds to ${\cal F}_3^{}$ and ${\cal F}_4^{}$;
the dash line corresponds to ${\cal F'}_3^{}$; the short dot line
corresponds to ${\cal \cal F'}_4^{}$; the dot horizontal line
corresponds to the experimental upper bound. }
\end{center}
\end{figure}
\begin{figure}
\begin{center}
\vspace{-0.5cm}
\includegraphics[]{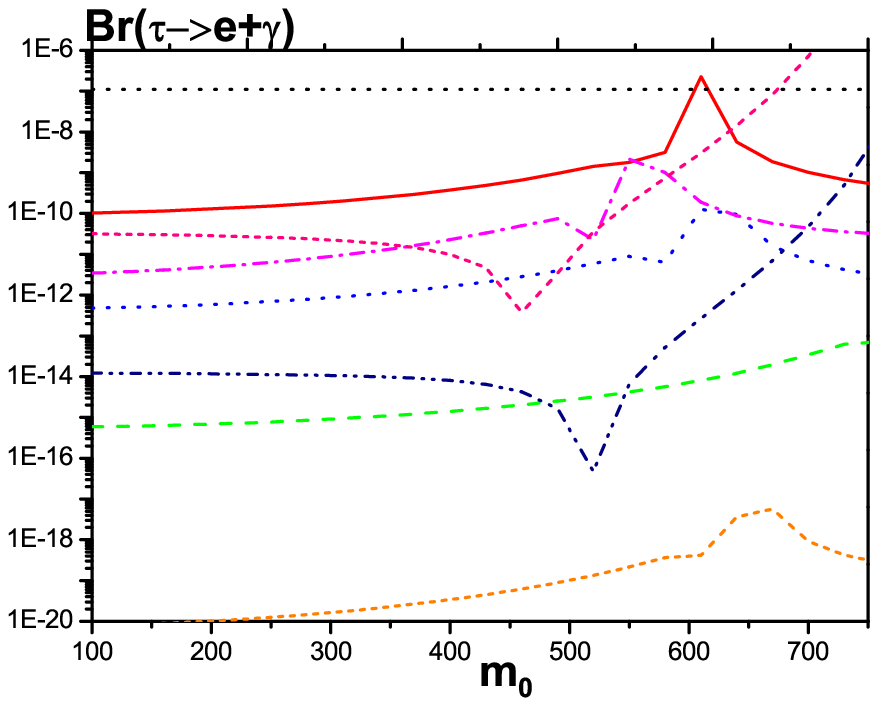} \caption{ Illustrative plot for ${\rm BR}(\tau\rightarrow e+\gamma)$
changing with $m^{}_{0}$. We take $\tan\beta=1.5$, $M_L^{}=1\ {
\rm Tev}$ and $M_R^{}=20\ {\rm Tev}$ in our plot. Here  the dash
line corresponds to ${\cal F}_1^{}$; the dot line corresponds to
${\cal F'}_1^{}$; the dash dot line correspond to ${\cal F}_2^{}$;
the dash dot dot line corresponds to ${\cal F'}_2^{}$; the solid
line corresponds to ${\cal F'}_3^{}$; the short dash line
corresponds to ${\cal F'}_4^{}$; the short dash line corresponds
to ${\cal F}_3^{}$ and ${\cal F}_4^{}$; the dot horizontal line
corresponds to the experimental upper bound.}
\end{center}
\end{figure}

\begin{thebibliography}{99}

\bibitem{exp1} Super-Kamiokande Collaboration,
Y. Fukuda {\it et al.}, Phys. Rev. Lett. {\bf 81}, 1562 (1998);
Phys. Rev. Lett. {\bf 86}, 5656 (2001).

\bibitem{exp2} SNO Collaboration, Q.R. Ahmad {\it et al.},
Phys. Rev. Lett. {\bf 87}, 071301 (2001); Phys. Rev. Lett. {\bf
89}, 011302 (2002).

\bibitem{exp3} KamLAND Collaboration, K. Eguchi {\it et al.},
Phys. Rev. Lett. {\bf 90}, 021802 (2003).

\bibitem{exp4} CHOOZ Collaboration, M. Apollonio {\it et al.},
Phys. Lett. B {\bf 420}, 397 (1998); Palo Verde Collaboration, F.
Boehm {\it et al.}, Phys. Rev. Lett. {\bf 84}, 3764 (2000).

\bibitem{exp5} K2K Collaboration, M.H. Ahn {\it et al.},
Phys. Rev. Lett. {\bf 90}, 041801 (2003).

\bibitem{Vissani}
A. Strumia and F. Vissani, hep-ph/0606054.

\bibitem{LR0} J.C. Pati and A. Salam, Phys. Rev. D {\bf 10}, 275
(1974); R.N. Mohapatra and J.C. Pati, Phys. Rev. D {\bf 11}, 566
(1975); Phys. Rev. D {\bf 11}, 2558 (1975); G. Senjanovic and R.N.
Mohapatra, Phys. Rev. D {\bf 12}, 1502 (1975).

\bibitem{Ji} See, e.g., R.N. Mohapatra and P.B. Pal, {\it Massive
Neutrinos in Physics and Astrophysics}, second edition (World
Scientific, 1998).

\bibitem{lrsusy1}
R.M. Fracis, M. Frank and C.S. Kalman, Phys. Rev. D {\bf 43}, 2369
(1990); R. Kuchimanchi and R.N. Mohapatra, Phys. Rev. D {\bf 48},
4352 (1993); Phys. Rev. Lett. {\bf 75}, 3989 (1995); C. Ailakh, A.
Melfo and G. Senjanovic, Nuovo Cim. A {\bf 110}, 615 (1997); C.
Aulakh, K. Benakli and G. Senjanovic, Phys. Rev. Lett. {\bf 79},
2188 (1997).

\bibitem{lrsusy2}
Z. Chacko and R.N. Mohapatra, Phys. Rev. D {\bf 58}, 015001
(1998); C.S. Aulakh, A. Melfo, A. Rasin and G. Senjanovic, Phys.
Rev. D {\bf 58}, 115007 (1998); B. Dutta and R.N. Mohapatra, Phys.
Rev. D {\bf 59}, 015018 (1999); M. Frank, H. Konig and M.
Pospelov, Eur. Phys. J. C {\bf 7}, 135 (1999).


\bibitem{origin0} R.N. Mohapatra and G. Senjanovic, Phys. Rev. Lett. {\bf 44}, 912
(1980); J. Schechterm and J.W.F. Valle, Phys. Rev. D {\bf 22},
2227 (1980); M. Magg and C. Wetterich, Phys. Lett. B {94}, 61
(1980); G. Lazarides, Q. Shafi and C. Wetterich, Nucl. Phys. B
{\bf 181}, 287 (1981).

\bibitem{first}
E.Kh. Akhmedov and M. Frogerio, Phys. Rev. Lett {\bf 96}, 061802
(2006); E.Kh. Akhmedov and M. Frigerio, JHEP, {\bf 0701}, 042
(2007); P. Hosteins, S. Lavignac and C.A. Savoy, Nucl. Phys. B
{\bf 755}, 137 (2006).

\bibitem{theotrical}
M. Frank, Phys. Rev. D {\bf 64}, 053013 (2001).

\bibitem{non-renormal}
J. Wess and B. Zumino, Phys. Lett. B {\bf 49}, 52 (1979); J.
Iliopoulos and B. Zumino, Nucl. Phys. B {\bf 76}, 310 (1974).

\bibitem{reduce}
W. Siegel, Phys. Lett. B {\bf 84}, 193 (1979); D.M. Capper, D.R.T.
Jones and P.V. Nieuwenhuizen, Nucl. Phys. B {167}, 479 (1980); S.
Antuch and M. Ratz, JHEP {\bf 0207}, 059 (2002).

\bibitem{technique}
P.H. Chankowski and Z. Pluciennik, Phys. Lett. B {\bf 316}, 312
(1993); K.S. Babu, C.N. Leung and J. Pantaleone, Phys. Lett. B
{\bf 319}, 191 (1993); S. Antusch, M. Drees, J. Kersten, M.
Lindner and M. Ratz, Phys. Lett. B {\bf 519}, 238 (2001); Phys.
Lett. B {\bf 525}, 130 (2002); S. Antusch, J. Kersten, M. Lindner,
M. Ratz and M.A. Schmidt, JHEP {\bf 0503}, 024 (2005); J.W. Mei,
Phys. Rev. D {\bf 71}, 073012 (2005); S. Luo, J.W. Mei and Z.Z.
Xing, Phys. Rev. D {\bf 72}, 053014 (2005); Z.Z. Xing, Phys. Lett.
B {\bf 633}, 550 (2006); Z.Z. Xing and H. Zhang, hep-ph/0601106;
W. Chao and H. Zhang, Phys. Rev. D {\bf 75}, 033003 (2007).

\bibitem{xianzhi}
A. Brignole and A. Rossi, Nucl. Phys. B {\bf 701}, 3 (2004);  F.R.
Joaquim and A. Rossi, Phys. Rev. Lett. {\bf 97}, 181801 (2006);
Nucl. Phys. B {\bf 765}, 71 (2007).

\bibitem{pdg}
Particle Data Group, W.M. Yao $et\ al$., J. Phys. G {\bf 33}, 1
(2006).


\bibitem{RGE}
N. Setzer and S. Spinner, Phys. Rev. D {\bf 71}, 115010 (2005).

\bibitem{lfv}
R. Barbieri, L.J. Hall and A. Strumia, Nucl. Phys. B {\bf 445},
219 (1995); J. Hisano, T. Moroi, K. Tobe and M. Yamaguchi, Phys.
Rev. D {\bf 53 }, 2442 (1996); J. Hisano and D. Nomura, Phys. Rev.
D {\bf 59}, 116005 (1999).

\bibitem{TB} P.F. Harrison, D.H. Perkins, and W.G. Scott, Phys.
Lett. B {\bf 530}, 167 (2002); Z.Z. Xing, Phys. Lett. B {\bf 533},
85 (2002); P.F. Harrison and W.G. Scott, Phys. Lett. B {\bf 535},
163 (2002); X.G. He and A. Zee, Phys. Lett. B {\bf 560}, 87
(2003).


\bibitem{gamma}
A. Joshipura, E.A. Paschos and W. Rodejohann, Nucl. Phys. B {\bf
611}, 227 (2001); JHEP {\bf 0108}, 029 (2001); W. Chao, S. Luo and
Z.Z. Xing, arXiv:0704.3838 [hep-ph].



\bibitem{new data}
MEGA Colloaboration, M.L. Brooks et al., Phys. Rev. Lett. {\bf
83}, 1521 (1999); BARBAR Collaboration, B. Aubert et al., Phys.
Rev. Lett. {\bf 96}, 041801 (2006).

\bibitem{future}
Super KEKB Physics Working Group, A.G. Akeroyd $et\
al$., hep-ex/0406071.

\bibitem{lfvlr}
R.N. Mohapatra, Z. Phys. C {\bf 56}, 117 (1992); M. Frank, Phys.
Rev. D {\bf 65}, 033011 (2002); K.S. Babu, B. Dutta and R.N.
Mohapatra, Phys. Rev. D {\bf 67}, 076006 (2003).

\bibitem{ncmixing}
V. Lyubimov, E.G. Novikov, V.Z. Nozik, E.F. Tretyakov and V.S.
Kosik, Phys. Lett. B {\bf 94}, 266 (1980).

\bibitem{neutralino}
M. Frank, C.S. Kalaman and H.N. Saif, Z. Phys. C {\bf 59 }655
(1993).



\end{thebibliography}
\end{document}